\documentclass[10pt]{article}
\usepackage{amsfonts}
\usepackage{stmaryrd}

\title{Contributions to the compositional semantics of
first-order predicate logic}
\author{Philip Kelly and M.H. van Emden\\
          \small Department of Computer Science\\
          \small University of Victoria, Canada}

\date{}
\begin{document}
\maketitle

\begin{abstract}
Henkin, Monk and Tarski gave a compositional semantics 
for first-order predicate logic.
We extend this work by including function symbols in the
language and by giving the denotation of the atomic formula
as a composition of the denotations of its predicate symbol
and of its tuple of arguments.
In addition we give the denotation of a term as a composition
of the denotations of its function symbol and of its tuple of
arguments.
\end{abstract}

\newtheorem{theorem}{Theorem}{}
\newtheorem{lemma}{Lemma}{}
\newtheorem{corollary}{Corollary}{}
\newtheorem{definition}{Definition}{}
\newtheorem{example}{Example}{}
\newtheorem{conjecture}{Conjecture}{}

\newcommand{\rel}[1]{\langle#1\rangle}
\newcommand{\ra}{\rightarrow}
\newcommand{\tr}{\triangleright}
\section{Introduction}

To introduce compositional semantics we can do no better
than to quote \cite{janssen01}:
\begin{quotation}
{\sl
The standard interpretation of compositionality is that for basic
expressions a meaning is given, and that operations are defined on
these meanings which yield meanings for compound expressions. Almost
all modern linguistic theories which give serious attention to
semantics follow this idea, [$\ldots$]
}
\end{quotation}
First-order predicate logic is one of the linguistic theories
where  serious attention is given to semantics.
As the language of this logic is formal,
its semantics should, {\sl a fortiori}, be compositional. 
The first semantics, \cite{tarski33}
(\cite{tarski35} in German, \cite{tarski56} in English)
was not compositional.
The first compositional semantics
(\cite{tarski52b}, \cite{hmt71})
omits function symbols from the language
and stops short of providing a compositional semantics
of atomic formulas and of terms.
In \cite{vnmdn06} this work was extended 
by including function symbols in the language.
It was also extended by giving
the denotation of the atomic formula
as a composition of the denotations of its predicate symbol
and of its tuple of arguments.
However, \cite{vnmdn06} was restricted to logic programs with Herbrand
semantics.
In the present paper we extend this work in several directions.
\begin{itemize} 
\item
We remove the restriction to Herbrand semantics and make the
results applicable to arbitrary interpretations.
\item
We remove the restriction to logic programs
\item
We not only provide a compositional semantics for atomic formulas,
but also for composite terms.
\end{itemize}

The initial impetus for the present paper was to give 
compositional semantics for terms and for atomic formulas
by relating the meaning of the function or predicate symbols
to the meanings of the tuples of the terms that occur as
arguments.
In pursuit of this purpose, interpretations acquire a central
position.
Once we obtained our basic results we needed to link these
to the compositional semantics already established in \cite{hmt71}.
The presence of interpretations and function symbols forced
us to deviate from their method.
The result is a compositional semantics for first-order
predicate logic that integrates our new results with a reformulation
of the pre-existing ones.
Although this reformulation is a minor departure from
\cite{hmt71}, we believe it has some interest in its own right.
\section{Sets, functions, tuples, and relations}\label{sec:sets}

In this section we establish the notation and terminology
for basic notions concerning sets, functions, tuples, and relations.

\subsection{Sets}
We assume the sets of standard (Zermelo-Fraenkel) set theory.
The sets specified in this paper owe their existence to the
axiom of specification:
$S'$ defined as equal to $\{x\in S\mid F\}$
where $F$ is an expression of which the truth depends of $x$.
In spite of the axiom of extensionality,
according to which a set is entirely determined by its elements,
we distinguish $S'$ from $\{x\in T\mid F\}$
even when $T\supseteq S$.

We denote the cardinality of a set $V$ by $|V|$.
For a finite $V$ with $|V| = n$, we freely confuse
the finite cardinals with the corresponding ordinals
and loosely refer to them as ``natural'' numbers.
As a result, locutions such as ``for all $i \in n$'' are common
as abbreviation of ``for all $i \in \{0, \ldots, n-1\}$''.

\subsection{Functions}
The set of all functions from set $S$ to set $T$,
not both empty\footnote{
Our reason for this restriction is that we want
$|S\to T|= |T|^{|S|}$ to hold,
and that this is defined except when both $S$ and $T$
are empty.
Thus $S$ may be empty.
In that case $S\ra T$ contains one function,
which we could call the empty function with target $T$.
Formally there is a different such function for every different $T$.
}, is denoted
$S \ra T$, so that we may write $f \in (S \ra T)$.
To relieve the overloaded term ``domain'' we call
$S$ the {\em source} and
$T$ the {\em target} of $S \ra T$ and of any $f$ belonging to it.
All functions have one argument.
But the source may be a set that is composed of other sets.

The value of $f$ at $x \in S$ is written as $fx$
or as $f_x$.
The process of ascertaining the value of $f$ at argument $x$
is called the {\sl application} of $f$ to $x$.
We often write ``$f(x)$'' instead of ``$fx$'';
this can help clarify what gets applied to what.

The composition $h$ of $f \in (S \ra T)$ and $g \in (T \ra U)$
is written as $g\circ f$ and is the function in $S \ra U$
defined by $x \mapsto g(f(x))$ for all $x \in S$.
It is often convenient to interchange the order of the
arguments in $g\circ f$.
In such situations we write $g\circ f$ as $f \tr g$.

\begin{definition}\label{def:setExt}
Given $f\in(S\to T)$ we also use $f$ to denote
the \underline{set extension} of $f$,
which is the function in $2^S\to 2^T$
with map $s \mapsto \{f(x) \mid x\in s\}$.\\
The \underline{inverse set extension} $f^{-1}$ of $f$ 
is the function in $2^T\to 2^S$
with map $t\mapsto\{x\in S \mid f(x)\in t\}$.
\end{definition}

\begin{lemma}\label{lem:unSymm}
Let $f$ be a function in $S\to T$.
We have\footnote{
That we do not have equality between the sets is necessitated
by the possibility that $f$ is not injective.
}
$s \subseteq f^{-1}(f(s))$ for all $s \subseteq S$.
We have $t = f(f^{-1}(t))$ for all $t \subseteq T$.
\end{lemma}

\begin{definition}\label{def:rangeRestr}
Let $f \in (S\to T)$ and let $S'$ be a subset of $S$.
The \emph{restriction} $f\downarrow S'$ of $f$ to $S'$
is the function in $S'\to T$ with map $x \mapsto f(x)$.
%
\hfill$\Box$
\end{definition}


\subsection{Tuples}

Tuples are regarded as functions.
When a function is a tuple, then the source of the function
is often referred to as its ``index set''.
The tuple $t \in (n \ra U)$, with the index set $n$ a natural number,
can be written as $(t_0,\ldots,t_{n-1})$.
Tuples can also be indexed by other sets.
For example, consider a certain tuple
$f \in (\{x,y,z\} \ra \{0,1\})$
specified by
$f(x) = 0$,
$f(y) = 1$, and
$f(z) = 0$.
We can achieve some abbreviation by listing the graph of $f$:
writing e.g.
$f = \{
\langle z,0 \rangle,
\langle x,0 \rangle,
\langle y,1 \rangle
\}$.
Another possibility is a tabular representation of this tuple:
$f =
\begin{tabular}{c|c|c}
$x$ & $y$ & $z$ \\
\hline
$0$ & $1$ & $0$
\end{tabular}
$.
This has the advantage of a compact representation
of a set of tuples of the same type:
$
\begin{tabular}{c|c|c}
$x$ & $y$ & $z$ \\
\hline
$0$ & $1$ & $0$\\
$1$ & $0$ & $1$
\end{tabular}
$.

The source $S$ of $f \in (S\to T)$ may be a set of tuples,
for example the set $n\to U$.
The value of $f$ at $t\in (n\to U)$
can be written $f(t_0,\ldots,t_{n-1})$ as alternative to $ft$.

\subsection{Relations}\label{sec:rel}

\begin{definition} \label{def:relation}
\begin{enumerate}
\item
A {\sl relation} $R$ is a triple
$\langle I,D,C \rangle$ of sets satisfying the constraint
that $C \subseteq (I\ra D)$.
$I \ra D$ is the \underline{domain} of the relation,
$I$ is the \underline{index set} of the relation,
and $C$ is the \underline{content} of the relation.
\item
When two relations have the same index set and the 
same domain, some of the common operations on and relations
between their contents can be extended:
\begin{tabbing}
MMMMMMMMMMMM\= MM\= \kill
$\rel{I,D,C_0} \subseteq \rel{I,D,C_1}$ \>iff\>
          $C_0 \subseteq C_1$\\
$\rel{I,D,C_0} \cap \rel{I,D,C_1}$ \>=\> $\rel{I,D,C_0\cap C_1}$\\
$\rel{I,D,C_0} \cup \rel{I,D,C_1}$ \>=\> $\rel{I,D,C_0\cup C_1}$\\
$\rel{I,D,C_0} \backslash \rel{I,D,C_1}$ \>=\>
   $\rel{I,D,C_0\backslash C_1}$\\
$\rel{I,D,C}^\sim $ \>=\> $\rel{I,D,(I\ra D)\backslash C}$
\end{tabbing}
\item
Let $R_0 = \rel{I_0,D,C_0}$ and $R_1 = \rel{I_1,D,C_1}$.
We write the \underline{bowtie} of $R_0$ and $R_1$
as $R_0 \Join R_1$
and define it to be $\rel{I_0\cup I_1,D,C}$
where 
$$
C = \{r\in((I_0\cup I_1)\to D) \mid 
    (r\downarrow I_0)\in R_0 \mbox{ and } (r\downarrow I_1)\in R_1\}.
$$
Similary for the \underline{oplus} of $R_0$ and $R_1$,
which is defined to be $\rel{I_0\cup I_1,D,C}$ where
$$
C = \{r\in((I_0\cup I_1)\to D) \mid 
    (r\downarrow I_0)\in R_0 \mbox{ or } (r\downarrow I_1)\in R_1\}.
$$
\item
Let $R=\rel{I,D,C}$ be a relation and 
let $I'$ be a subset of $I$.
Let $R'=\rel{I',D,C'}$ be a relation.
We write the \underline{projection} of $R$ on $I'$ as $\pi_{I'}(R)$
and define it to be
$$\rel{I',D,
  \{c\downarrow I' \mid c\in C\}
}.$$
We write the \underline{cylinder} in $I$ on $R'$ as $\rho_{I}(R')$
and define it to be
$$ \rel
{I,D,
 \{c\in(I\to D) \mid (c\downarrow I')\in C'\}
}.$$

\end{enumerate}
\hfill$\Box$
\end{definition}
When referring to a relation,
its index set and domain are often clear from the context.
In such cases we refer to the relation by its content only.

\begin{example}\label{ex:miscRel0}
Let $I_0 = \{x,y\}$, $I_1 = \{y,z\}$, and $D = \{a,b\}$,
let $R_0=
\rel{I_0,D,
\begin{tabular}{c|c}
$x$ & $y$ \\
\hline
$a$ & $b$ \\
$b$ & $a$
\end{tabular}
}
$
and let $R_1 =
\rel{I_1,D,
\begin{tabular}{c|c}
$y$ & $z$ \\
\hline
$a$ & $a$ \\
$b$ & $a$
\end{tabular}
}
.
$

We have
$R_0\Join R_1 =
\rel{\{x,y,z\},D,
\begin{tabular}{c|c|c}
$x$ & $y$ & $z$ \\
\hline
$a$ & $b$ & $a$ \\ 
$b$ & $a$ & $a$
\end{tabular}
}
.
$
We have
$R_0\; \oplus\; R_1 =
\rel{\{x,y,z\},D,
\begin{tabular}{c|c|c}
$x$ & $y$ & $z$ \\
\hline
$a$ & $b$ & $a$ \\ 
$a$ & $b$ & $b$ \\ 
$b$ & $a$ & $a$ \\ 
$b$ & $a$ & $b$ \\ 
$a$ & $a$ & $a$ \\ 
$b$ & $b$ & $a$
\end{tabular}
}
.
$
\end{example}

\begin{example}\label{ex:miscRel}
$$
\pi_{\{y\}}(
\rel{\{x,y\},\{a,b\},
\begin{tabular}{c|c}
$x$ & $y$ \\
\hline
$a$ & $a$ \\
$a$ & $b$ \\
$b$ & $b$
\end{tabular}
\;
}
)
=
\rel{\{y\},\{a,b\},
\begin{tabular}{c}
$y$ \\
\hline
$a$ \\
$b$
\end{tabular}
\;
}
.
$$

$$
\rho_{\{x,y\}}(
\rel{\{y\},\{a,b\},
\begin{tabular}{c}
$y$ \\
\hline
$a$ \\
$b$ 
\end{tabular}
\;
})
=
\rel{\{x,y\},\{a,b\},
\begin{tabular}{c|c}
$x$ & $y$ \\
\hline
$a$ & $a$ \\
$a$ & $b$ \\
$b$ & $a$ \\
$b$ & $b$
\end{tabular}
\;
}
.
$$

\end{example}


$\Join$ is similar to the numerous variants of ``join'' in
relational databases (natural join, equijoin, crossjoin, thetajoin,
and perhaps others). To prevent confusion, we refer to the
$\Join$ defined here as ``bow tie''.
$\oplus$ is forced into existence as the counterpart
of bow tie, and is new to us.
For lack of a better term, we refer to it as ``oplus''.
``Projection'' and ``cylinder'' are inspired by
Tarski {\sl et al.} \cite{hmt71}.

Note that bow tie and oplus are generalizations of intersection
and union.
In the cylindric set algebra of \cite{hmt71} all relations have
the same type (with the same infinite index set)
so that the generalization represented by bow tie
and oplus is not needed.

If the index set $I$ is $n$ for some ordinal $n$,
then the relation is said to be an $n$-ary relation.
It is a subset of $D^n$.
It is said to be a type-P relation because it can be
the interpretation of a predicate symbol.

If $I$ is the set of free variables of an expression,
then the relation is said to be {\sl type C}.
Thus the relation defined in Definition~\ref{def:denotF}
on page \pageref{def:denotF} is a type-C relation.
Here ``C'' is from ``cylinder'' because the cylinders in
cylindric algebra \cite{hmt71} are
relations of this kind.
Type-C relations cannot be interpretations of predicate
symbols, as the arguments in atomic formulas are indexed
numerically.

\begin{lemma}
Let
$R_0 = \rel{I_0,D,C_0}$,
$R_1 = \rel{I_1,D,C_1}$,
and $I = I_0 \cup I_1$.
We have
\begin{eqnarray*}
R_0 \Join R_1 &=& \rho_I(R_0) \cap \rho_I(R_1)   \\
R_0 \oplus R_1 &=& \rho_I(R_0) \cup \rho_I(R_1)   \\
\end{eqnarray*}
\end{lemma}
{\sl Proof}\\
Immediate from the definitions.
\hfill$\Box$

\begin{lemma}
\begin{eqnarray*}
\rel{I',D,C} &=& \pi_{I'}(\rho_I(\rel{I',D,C})) \\
\rel{I,D,C} &\subseteq& \rho_I(\pi_{I'}(\rel{I,D,C}))
\end{eqnarray*}
\hfill$\Box$
\end{lemma}
{\sl Proof}\\
Note that the $\pi_{I'}$ and $\rho_I(\pi_{I}$
of Definition~\ref{def:relation} are respectively
a set extension and inverse set extension
(Definition~\ref{def:setExt}) of function restriction
(Definition~\ref{def:rangeRestr}).
With this identification, the Lemma to be proved
can be seen to follow from Lemma~\ref{lem:unSymm}.
\hfill$\Box$

We will encounter a situation involving a set
$X$ of $n$ variables, a set $D$, and a tuple $\alpha$
of type $X \ra D$.
Let $x = (x_0,\ldots,x_{n-1})$,
a tuple of type $n \ra X$,
be an enumeration of $X$.
That is, there is no repeated occurrence
of any variable in $x = (x_0,\ldots,x_{n-1})$.
The tuple $t$ defined as $x \tr \alpha$ is of type $n \ra D$.
We extend the function composition $\tr$ from individual
tuples to relations, that is, to sets of tuples of the same type.

\begin{definition}\label{def:relComp}
$$
\rel{S,T,C_0} \tr \rel{T,U,C_1}
= 
\rel{S,U,\{d_0\tr d_1 \mid d_0\in C_0 \wedge d_1\in C_1\}}.
$$
With $d_0\in(S\ra T)$ and $d_1\in(T\ra U)$
we write $\rel{S,T,C_0} \tr d_1$ for
$\rel{S,T,C_0} \tr \rel{T,U,\{d_1\}}$
and
$d_0\tr\rel{T,U,C_1}$ for
$\rel{S,T,\{d_0\}} \tr \rel{T,U,C_1}$.
\hfill$\Box$
\end{definition}

\begin{lemma}\label{lem:relCompProp}
Let $S$ and $T$ be sets, not both empty.
We have
\begin{eqnarray*}
S \to S &=& (S\to T)\tr(T\to S)   \\
S \to T &=& (S\to S)\tr(S\to T) = (S\to T)\tr(T\to T)
\end{eqnarray*}
\hfill$\Box$
\end{lemma}
\section{Satisfaction semantics}\label{sec:satSem}

One aspect of semantics defines the conditions under which a sentence,
i.e. a {\sl closed} formula, is true in (is {\sl satisfied} by)
an interpretation.
We first review this aspect of semantics,
which we call {\sl satisfaction semantics}. 
Then we show how it can also be applied to the definition
of relations and functions.

\subsection{Logic preliminaries}

\begin{definition}\label{def:sigEtc}
\begin{enumerate}
\item A \underline{signature} consists of
\begin{enumerate}
\item
A set of constant symbols.
\item
Sets of $n$-ary predicate symbols for
nonnegative integers $n$.
We denote the arity of a predicate symbol $q$ by $|q|$.
\item
Sets of $n$-ary function symbols for
positive integers $n$.
We denote the arity of a function symbol $f$ by $|f|$.
\end{enumerate}
A term (formula) of first-order predicate logic is an
$L$-term ($L$-formula) if the function and predicate symbols
have the names and arities specified in $L$.

\item A \underline{structure} consists of a universe $D$
(also referred to as ``domain''), which is a set,
and numerically-indexed relations and functions
over $D$.\\[4mm]
For a given signature $L$, a structure $S$ with universe $D$
is an $L$-structure whenever
\begin{enumerate}
\item
each constant in $L$ is associated with an element of $D$,
\item
each predicate symbol $p$ in $L$ is associated with a relation
in $S$ of type $D^{|p|}$,
\item
and
each function symbol $f$ in $L$ is associated with a function
in $S$ of type $D^{|f|} \rightarrow D$.
\end{enumerate}
\item
Let a signature $L$ and an $L$-structure $S$
with domain $D$ be given.
An \underline{interpretation} of an $L$-term $T$ consists of
a mapping from each function symbol $f$ of $T$
to an $|f|$-ary function of $S$.
An interpretation of an $L$-formula $F$ consists of
a mapping from each predicate symbol $p$ of $F$
to a $|p|$-ary relation of $S$.\\
Note that $I(f)$ and $I(p)$ are numerically indexed.
\end{enumerate}
\end{definition}

\subsection{Satisfaction}
Given a signature $L$ and an interpretation $I$
for $L$-formulas. 
Let $D$ be the universe of discourse of $I$.
The truth value of an $L$-formula $F$ with $V$ as set
of free variables depends on $I$
for the interpretation of the predicate and function symbols.
It also depends on an assignment $\alpha$ of individuals in $D$
to the variables in $V$.
That is, $\alpha$ is a function of type $V \to D$.
With these dependencies in mind,
we write $M^I_\alpha(F)$ for the meaning,
that is a truth value, of formula $F$.

$M^I_\alpha(F)$ is an expression in the metalanguage,
which is informal mathematics.
Here $F$ is the metalanguage name for a formula in
first-order predicate logic.

The meaning of $M^I_\alpha(F)$, with $F$ a formula,
is a truth value.
The meaning of $M^I_\alpha(t)$, with $t$ a term,
is an individual in the universe of discourse $D$.
The meaning of $M^I_\alpha(t_0,\ldots,t_{n-1})$
is an $n$-tuple of individuals in the universe of discourse $D$.

Several well-known textbooks
\cite{mendelson64,shoenfield67,enderton72,grzegorczyk74}
define satisfaction semantics in substantial agreement;
\cite{mendelson64,grzegorczyk74} attribute the definition
to Tarski \cite{tarski33}, which has been translated in \cite{tarski35}
and in \cite{tarski56}.
Tarski's paper addresses philosophers and argues that
the concept of truth can only be defined without danger of
paradox in formalized languages and then only in a suitable
class of these.
By the 1960's Tarski's satisfaction semantics
had become folklore to a sufficient extent that
Shoenfield \cite{shoenfield67} and Enderton \cite{enderton72}
give it without attribution.

\begin{definition}\label{def:textbook}
Let $L$ be a signature
including at least one constant, a function symbol $f$,
and a predicate symbol $p$.
Let $I$ be an interpretation for $L$
with $D$ as universe of discourse.

\begin{enumerate}
\item
$M^I_\alpha(t) = I(t)$ if $t$ is a constant.
\item
$M^I_\alpha(t) = \alpha(t)$ if $t$ is a variable.
\item
$M^I_\alpha(t) =
(I(f))(M^I_\alpha(t_0),\ldots,M^I_\alpha(t_{n-1}))$
 if $t$ is $f(t_0,\ldots,t_{n-1})$.
\item
If $p$ is a predicate symbol, then
$M^I_\alpha(p(t_0,\ldots,t_{n-1}))$ is true iff
$$(M^I_\alpha(t_0),\ldots,M^I_\alpha(t_{n-1}))
     \in I(p).$$
\item
$M^I_\alpha(\neg F)$ is true iff not $M^I_\alpha(F)$.
\item
Let $F_0$ have $V_0$ as set of free variables,
let $F_1$ have $V_1$ as set of free variables,
and let $V = V_0 \cup V_1$.
Let $\alpha \in V\to D$.\\
We define
$M^I_\alpha(F_0 \wedge F_1)$ iff
$M^I_{\alpha\downarrow V_0}(F_0)$
and $M^I_{\alpha\downarrow V_1}(F_1)$.\\
We define
$M^I_\alpha(F_0 \vee F_1)$ iff
$M^I_{\alpha\downarrow V_0}(F_0)$
or $M^I_{\alpha\downarrow V_1}(F_1)$.
\item\label{item:quant}
Let $F$ be a formula with set $V$ of free variables.
Let $\alpha$ be a tuple of type $V\ra D$.
Let, for some $\{x_0,\ldots,x_{n-1}\}\subseteq V$
and some $d_i\in D$ for $i\in n$,
$\alpha'\in (V\ra D)$ be such that 
$\alpha'(x_i) = d_i$ for all $i\in n$ and $\alpha'(v)=\alpha(v)$
if $v\in V\backslash \{x_0,\ldots,x_{n-1}\}$.

We define
$M^I_{\alpha\downarrow(V\backslash\{x_0,\ldots,x_{n-1}\})}
    (\exists x_0\ldots\exists x_{n-1}.\; F)$ iff
there exist $d_0,\ldots,d_{n-1}\in D$ such that $M^I_{\alpha'}(F)$.

We define
$M^I_{\alpha\downarrow(V\backslash\{x_0,\ldots,x_{n-1}\})}
    (\forall x_0\ldots\forall x_{n-1}.\; F)$ iff
for all $d_0,\ldots,d_{n-1}\in D$ it is the case that $M^I_{\alpha'}(F)$.

\end{enumerate}

Conventionally, the language of first-order predicate logic
has terms and formulas as its syntactic categories.
To these we add $n$-tuples of terms.
Their denotation is defined as follows.

$$M^I_\alpha(t_0,\ldots,t_{n-1}) =
(M^I_\alpha(t_0),\ldots,M^I_\alpha(t_{n-1}))\in D^n.$$
\hfill$\Box$
\end{definition}
We may write $\exists V'.F$
instead of 
$\exists x_0\ldots\exists x_{n-1}.\; F$
when $V'=\{x_0,\ldots,x_{n-1}\}$ 
is a subset of the set of free variables of $F$.
Similarly for universal quantification.

\begin{lemma}\label{lem:textBtuple}
In the context of Definition~\ref{def:textbook} we
have
$$
M_\alpha^I(f(t_0,\ldots,t_{n-1})) =
(I(f))M^I_\alpha(t_0,\ldots,t_{n-1}).
$$
\hfill$\Box$
\end{lemma}
{\sl Proof}\\
Apply Definition~\ref{def:textbook}.
\hfill$\Box$

\subsection{Application of satisfaction semantics
    to the definition of relations}

In his treatment of set theory,
Suppes \cite{suppes60} gives the Axiom of Separation
(Zermelo's ``Aussonderungsaxiom'') as
\begin{equation}\label{eq:sep0}
\exists y\forall x\;[x\in y \leftrightarrow
                     (x\in z \wedge \varphi(x))]
\end{equation}
where $z$ is a given set.
This can be paraphrased to
\begin{equation}\label{eq:sep1}
\mbox{if } z \mbox{ is a set, then }
\{x\in z\mid\varphi(x)\}
\mbox{ is a set.}
\end{equation}

In this general formulation it is not specified in what
language $\varphi(x)$ is expressed.
We make $\varphi(x)$ more specific by expressing it as
a formula $F$ of
first-order predicate logic.
Let us call $V$ the set of free variables in $F$.
Then, given an interpretation $I$,
the meaning of $F$ is determined by an assignment $\alpha$
of elements of $D$ to the variables in $V$.
That is, in (\ref{eq:sep1}) the given set $z$ becomes
$V\ra D$ and we get for the subset of $z$ called into
existence by the Axiom of Separation:
\begin{equation}\label{eq:sep2}
\{\alpha\in(V\ra D) \mid F \mbox{ is true in } I
      \mbox{ with } \alpha\},
\end{equation}
which is, using the formalism developed in this section,
\begin{equation}\label{eq:sep3}
\{\alpha\in(V\ra D) \mid M_\alpha^I(F)\} 
\end{equation}

We arrived at (\ref{eq:sep3}) in an attempt to clarify
the Axiom of Separation by means of first-order predicate logic.
This expression can also be used to define the meaning of
a formula when an interpretation is given,
abstracting away from the assignment $\alpha$.

\begin{definition}\label{def:denotF}
The meaning of a formula $F$ with set $V$ of free variables,
given interpretation $I$ with domain $D$ is written as
$M^I(F)$ and defined to be
$\{\alpha\in(V\ra D) \mid M_\alpha^I(F)\}.$ 
\end{definition}

Definition~\ref{def:denotF} defines the meaning of $F$ as
a subset of $V\to D$.
That is, as a set of tuples, which is the usual definition of
a {\sl relation}.
A compositional semantics of first-order predicate logic
investigates how the meanings of the subformulas of $F$ are connected to
the meaning of $F$ itself.
As these meanings are now seen to be relations,
we need a suitable set of operations on relations.
These were defined in Section~\ref{sec:rel}.

\subsection{Application of satisfaction semantics
    to the definition of functions}

Tarski {\sl et al.} \cite{tarski52b,hmt71}
defined certain relations as denotations
of formulas with free variables.
A natural counterpart would be to define functions as denotations
of terms.
However, this was omitted from \cite{tarski52b,hmt71}, as the language of logic
there does not include function symbols.
In this section we introduce a counterpart of Definition~\ref{def:denotF}
that defines a function as denotation of any term.
Such a definition is only interesting if the denotation of a composite
term is related in a plausible way to the denotations of its constituent
terms.
This we do in Section~\ref{sec:decTerm}.

Let $t$ be a term with set $V$ of variables.
Let $I$ be an interpretation for $t$ with universe of discourse $D$.
Every $\alpha \in (V\to D)$ determines $M_\alpha^I(t) \in D$.
In other words, $I$ and $t$ determine a function
of type $(V\to D)\to D$.
Hence we define the meaning $M^I(t) \in ((V\to D)\to D)$ of $t$
as a function in the following way.
\begin{definition}\label{def:denotT}
Given an interpretation $I$ with domain $D$.
Let $t$ be a term with set $V$ of variables.
We write the meaning of $t$ as $M^I(t)$ and define it as
the function in $(V\to D)\to D$ with map
$
\alpha \mapsto M^I_\alpha(t).
$
\hfill$\Box$
\end{definition}
\section{Compositional semantics of composite formulas}

\begin{lemma}\label{lem:MalphaAlt}
$M_\alpha^I(F)$ iff 
$\alpha\in M^I(F)$ iff 
$\alpha\in \{\beta\in(V\to D)\mid M_\beta^I(F)\}$.
\end{lemma}
\begin{definition}\label{def:denotTT}
Let $t_0,\ldots,t_{n-1}$ be a tuple of terms with set $V$
of variables.
Let $I$ be an interpretation for $t_0,\ldots,t_{n-1}$
with universe $D$.
We define
$$ M^I(t_0,\ldots,t_{n-1}) =
\{M_\alpha^I(t_0,\ldots,t_{n-1})
   \mid \alpha \in (V\to D)\}.
$$
\end{definition}
\hfill$\Box$

We provide examples and lemmas to build intuition about
the consequences of Definitions~\ref{def:denotF} and
\ref{def:denotTT}.
\begin{example}\label{ex:diag}
In Definition~\ref{def:denotTT} one may regard
the tuples of terms as a language  to specify subsets of
$n\to D$, alias relations of type $n\to D$.
For example $M^I(x,\ldots,x)$ should denote the diagonal of
the set diagonals in $n\to D$.
\begin{tabbing}
MMMMMMMMMMMMMMMMMMMM\= MM\= \kill
$M^I(x,\ldots,x)$ \> $=$ \>(1)  \\
$\{M_\alpha^I(x,\ldots,x)
  \mid \alpha \in (\{x\}\to D)\}$ \> $=$ \>(2)  \\
$\{(M_\alpha^I(x),\ldots,M_\alpha^I(x))
    \mid \alpha \in (\{x\}\to D)\}$ \> $=$ \>(3)  \\
$\{(\alpha(x),\ldots,\alpha(x))
    \mid \alpha \in (\{x\}\to D)\}$ \> $=$ \>(4)  \\
$\{(d,\ldots,d) \mid d \in D\}$.
\end{tabbing}
(1) Definition~\ref{def:denotTT};
(2) Lemma~\ref{lem:textBtuple};
(3) Definition~\ref{def:textbook};
(4) let $d=\alpha(x)$ and note that for every $d\in D$
there is exactly on $\alpha \in (\{x\}\to D)$
such that $\alpha(x)=d$.
\hfill$\Box$
\end{example}

\begin{example}\label{ex:trivCase}
Let $I$ be an interpretation with universe of discourse $D$.
Let $\{x_0,\ldots,x_{n-1}\}$ be an enumeration
of a set $V$ of variables.
In $M^I(x_0,\ldots,x_{n-1}) \subseteq (n\to D)$
the arguments are at their least constrained,
so this expression should denote all of $n\to D.$
\begin{tabbing}
MMMMMMMMMMMMMMMMMMMMMMMMMMMMMMMMMMMM\= MM\= \kill
$M^I(x_0,\ldots,x_{n-1})$ \> $=$ \> (1) \\
$\{(M_\alpha^I(x_0),\ldots,M_\alpha^I(x_{n-1}))
    \mid \alpha \in (V\to D)\}$ \> $=$ \> (2) \\
$\{(\alpha(x_0),\ldots,\alpha(x_{n-1}))
    \mid \alpha \in (V\to D)\}$ \> $=$ \> (3) \\
$\{(\alpha(x_0),\ldots,\alpha(x_{n-1}))
    \mid (x_0,\ldots,x_{n-1})\tr \alpha \in
       ((x_0,\ldots,x_{n-1})\tr (V\to D))\}$ \> $=$ \> (4) \\
$\{(d_0,\ldots,d_{n-1})
    \mid (x_0,\ldots,x_{n-1})\tr \alpha \in
       ((x_0,\ldots,x_{n-1})\tr (V\to D))\}$ \> $=$ \> (5) \\
$\{(d_0,\ldots,d_{n-1})
    \mid (x_0,\ldots,x_{n-1})\tr \alpha \in
       (n\to D))\}$ \> $=$ \> (6) \\
$\{(d_0,\ldots,d_{n-1})
    \mid (d_0,\ldots,d_{n-1})\in (n\to D)\}$ \> $=$ \> (7) \\
$n\to D$.
\end{tabbing}
(1) Definition~\ref{def:denotTT} and Lemma~\ref{lem:textBtuple};
(2) Definition~\ref{def:textbook};
(3) rewrite condition;
(4) introduce the names $d_i$ for $\alpha(x_i)$;
(5) Definition~\ref{def:relComp};
(6) definition of function composition;
(7) meaning of set comprehension.
\hfill$\Box$
\end{example}

The compositional semantics of conjunction and disjunction is given
by Theorem~\ref{thm:compConj}.
\begin{theorem}\label{thm:compConj}
Let $F_0$ and $F_1$ be formulas.
Let $I$ be an interpretation for the predicate and function
symbols of $F_0$ and $F_1$.
We have\\
(a)
$ M^I(F_0 \wedge F_1) = M^I(F_0) \Join M^I(F_0)$\\
(b) $ M^I(F_0 \vee F_1) = M^I(F_0) \oplus M^I(F_0)$ and\\
(c) $M^I(\neg F) = M^I(F)^\sim$.
\end{theorem}
{\sl Proof}\\
Let $V_0$ and $V_1$ be the sets of free variables of $F_0$ and $F_1$,
respectively.
Let $D$ be the domain of $I$.
\begin{tabbing}
MMMMMMMMMMMMMMMMMMMMMMMMMMMMMMMMMMMMMMMMM\= MM\= \kill
$ M^I(F_0 \wedge F_1)$ \> = \> (1) \\
$\{\alpha \in ((V_0\cup V_1)\to D) \mid
    M_\alpha^I(F_0 \wedge F_1)\}$ \> = \> (2) \\
$\{\alpha \in ((V_0\cup V_1)\to D) \mid
    M_{\alpha\downarrow V_0}^I(F_0)
      \mbox{ and }
          M_{\alpha\downarrow V_1}^I(F_1) \}$ \> = \> (3) \\
$\{\alpha \in ((V_0\cup V_1)\to D) \mid
    \alpha\downarrow V_0\in
        \{\beta\in V_0\mid M_\beta^I(F_0)\}
      \mbox{ and }
    \alpha\downarrow V_1\in
        \{\gamma\in V_1\mid M_\gamma^I(F_1)\}\}$
                                             \> = \> (4) \\
$\{\alpha \in ((V_0\cup V_1)\to D) \mid
    \alpha\downarrow V_0\in M^I(F_0)
      \mbox{ and }
    \alpha\downarrow V_1\in M^I(F_1)\}$       \> = \> (5) \\
$M^I(F_0) \Join M^I(F_1).$
\end{tabbing}
(1): Definition~\ref{def:denotF},
(2): Definition~\ref{def:textbook},
(3): Lemma~\ref{lem:MalphaAlt},
(4): meaning of set comprehension, and
(5): Definition~\ref{def:relation}.

There is a similar proof of part (b).
Proof of (c):
\begin{tabbing}
MMMMMMMMMMMMMMMMMM\= MM\= \kill
$ M^I(\neg F)$ \> = \> (1) \\
$ \{\alpha \in (V\to D) \mid M^I_\alpha(\neg F)\}$ \> = \> (2) \\
$ \{\alpha \in (V\to D) \mid
    \mbox{ not } M^I_\alpha(F)\}$ \> = \> (3) \\
$ (V\to D)\backslash \{\alpha \in (V\to D) \mid
       M^I_\alpha(F)\}$ \> = \> (4) \\
$(V\to D)\backslash M^I(F).$
\end{tabbing}
(1) Definition~\ref{def:denotF},
(2) Definition~\ref{def:textbook},
(3) meaning of set comprehension,
(4) Definition~\ref{def:denotF}.

\hfill$\Box$

This is basically a result from \cite{hmt71}, where conjunction
corresponds to intersection of relations.
The difference arises because there all relations have the same
index set; in that special case bow tie and intersection coincide,
as do oplus and union.

%
%

\begin{theorem}\label{thm:compEx}
Let $F$ be a formula with $V$ as set of free variables.
Let $I$ be an interpretation for the predicate and function
symbols with $D$ as universe of discourse.
Let $V'$ be a subset of V.
We have $M^I(\exists V'.\; F) = \pi_{V\backslash V'}M^I(F)$.
\end{theorem}
{\sl Proof}
\begin{tabbing}
MMMMMMMMMMMMMMMMMMMMMMMMMMMMMMMMMMMMM\= MM\= \kill
$ M^I(\exists V'.\; F)$ \> = \> (1) \\
$ \{\alpha'\in((V\backslash V')\ra D) \mid
     M_{\alpha'}^I(\exists V'.\; F)\}$ \> = \> (2) \\
$ \{\alpha'\in((V\backslash V')\ra D) \mid
  \exists\alpha\in(V\to D)
    \mbox{ s.t. } M^I_{\alpha}(F) \mbox{ and }
     \alpha\downarrow(V\backslash V') = \alpha' \}$ \> = \> (3) \\
$ \{\alpha\downarrow(V\backslash V') \mid \alpha\in M^I(F)\}$ 
\end{tabbing}
(1) Definition~\ref{def:denotF},
(2) Definition~\ref{def:textbook}, and
(3) Definition~\ref{def:denotF}.

This completes the proof, as the final expression is the content
of the relation $\pi_{V\backslash V'}M^I(F)$ according to 
Definition~\ref{def:relation}.
\hfill$\Box$

\section{Compositional semantics of atomic formulas}
Tarski {\sl et al.} \cite{tarski52b,hmt71} introduced
denotations for logic formulas only as motivation
for the introduction of cylindric algebras.
They did not consider decomposition of atomic formulas.
Our primary interest is compositional semantics.
This point of view demands a semantic analysis
of the inner structure of atomic formulas.

$M^I(p(t_0,\ldots,t_{n-1}))$ is determined
by Definition~\ref{def:denotF}.
Yet it is also 
determined by the denotation of its constituent $p$
(according to $I$) and by
that of its constituent $(t_0,\ldots,t_{n-1})$
(according to Definition~\ref{def:denotTT}).
The following theorem shows how
$M^I(p(t_0,\ldots,t_{n-1}))$ is determined
by the denotations of its constituents.

\begin{theorem}\label{thm:atomCompB}
Let $L$ be a signature with a predicate symbol $p$.
Let $I$ be an $L$-interpretation with universe of discourse $D$.
Let $t_0,\ldots,t_{n-1}$ be $L$-terms with $V$
as union of the sets of their variables.

We have
$$ M^I(p(t_0,\ldots,t_{n-1})) =
(V\to n)\tr (I(p) \cap M^I(t_0,\ldots,t_{n-1})).
$$ 
\end{theorem}
{\sl Proof}
(1) Lemma~\ref{lem:relCompProp},
\begin{tabbing}
MMMMMMMMMMMMMMMMMMMMMMMMMMMMMMMMMMMMM\= MM\= \kill
$(V\to n) \tr (I(p) \cap M^I(t_0,\ldots,t_{n-1}))$ \> $=$
  \> (1) \\
$(V\to n) \tr (I(p) \cap
  \{M_\alpha^I(t_0,\ldots,t_{n-1})\mid \alpha\in(V\to D)\}$ \> $=$
  \> (2) \\
$(V\to n) \tr (I(p) \cap
  \{d\in(n\to D)\mid \exists \alpha\in(V\to D).\;
     d = M_\alpha^I(t_0,\ldots,t_{n-1}) \}$ \> $=$
  \> (3) \\
$(V\to n) \tr \{d\in(n\to D)\mid \exists \alpha\in(V\to D).\;
     d\in I(p) \wedge d = M_\alpha^I(t_0,\ldots,t_{n-1}) \}$ \> $=$
  \> (4) \\
$(V\to n) \tr \{d\in(n\to D)\mid \exists \alpha\in(V\to D).\;
     M_\alpha^I(p(t_0,\ldots,t_{n-1})) \}$ \> $=$
  \> (5) \\
$\{\gamma\tr d \mid \gamma\in(V\to n) \wedge 
  d\in(n\to D)\wedge \exists \alpha\in(V\to D) \wedge
     M_\alpha^I(p(t_0,\ldots,t_{n-1})) \}$ \> $=$
  \> (6) \\
$\{\alpha\in(V\to D) \mid 
     M_\alpha^I(p(t_0,\ldots,t_{n-1})) \}$ \> $=$
  \> (7) \\
$M^I(p(t_0,\ldots,t_{n-1})).$
\end{tabbing}
(1) Definition~\ref{def:denotTT}\\ 
(2) note that
the set comprehension expression is the range of the function\\
$\lambda\alpha\in(V\to D).\;M_\alpha^I(t_0,\ldots,t_{n-1})$
and use the fact that\\
$\{f(x) \mid x\in S\}
=
\{y\in T \mid \exists x\in S.\;y=f(x)\}
$\\
(3)
eliminate the set intersection with $I(p)$ by inserting the equivalent
condition $d\in I(p)$\\
(4)
Definition~\ref{def:textbook}\\
(5)
Definition~\ref{def:relComp}\\
(6)
give $\gamma\tr d$ the name $\alpha'$;
this eliminates $\gamma$ and $d$,
and identify $\alpha'$ with the $\alpha$ that exists
according to $\exists \alpha\in(V\to D)$ in the condition\\
(7) Definition~\ref{def:denotF}.
\hfill$\Box$

\begin{theorem}\label{thm:atomComp}
Let $L$ be a signature containing a predicate symbol $p$.
Let $I$ be an $L$-interpretation with universe of discourse $D$.
Let $t_0,\ldots,t_{n-1}$ be $L$-terms with $V$ as union of their
sets of variables.

We have
$$
(n\to V) \tr M^I(p(t_0,\ldots,t_{n-1}))
=
I(p) \cap M^I(t_0,\ldots,t_{n-1}).
$$ 
\end{theorem}
{\sl Proof}
\begin{tabbing}
MMMMMMMMMMMMMMMMMMMMMMMMMMMMMMMMMMMMM\= MM\= \kill
$I(p) \cap M^I(t_0,\ldots,t_{n-1})$ \> $=$ \> (1) \\
$(n\to n) \tr (I(p) \cap M^I(t_0,\ldots,t_{n-1}))$ \> $=$ \> (2) \\
$((n\to V) \tr (V\to n)) \tr
  (I(p) \cap M^I(t_0,\ldots,t_{n-1}))$ \> $=$ \> (3) \\
$(n\to V) \tr ((V\to n)) \tr
  (I(p) \cap M^I(t_0,\ldots,t_{n-1})))$ \> $=$ \> (4) \\
$(n\to V) \tr M^I(p(t_0,\ldots,t_{n-1}))$.
\end{tabbing}
(1): Lemma~\ref{lem:relCompProp},
(2): Lemma~\ref{lem:relCompProp},
(3): Lemma~\ref{lem:relCompProp},
(4): Theorem~\ref{thm:atomCompB}.
\hfill$\Box$

\begin{corollary}\label{cor:free}
Let $L$ be a signature including predicate symbol $p$.
Let $I$ be an $L$-interpretation with $D$ as universe of
discourse.
Let $\{x_0,\ldots,x_{n-1}\}$ be a set of $n$ variables.
We have 
$$
(n\to \{x_0,\ldots,x_{n-1}\}) \tr M^I(p(x_0,\ldots,x_{n-1})) = I(p).
$$
\end{corollary}

{\sl Proof}
\begin{tabbing}
MMMMMMMMMMMMMMMMMMMMMM\= MM\= \kill
$(n\to \{x_0,\ldots,x_{n-1}\}) \tr M^I(p(x_0,\ldots,x_{n-1}))$
  \> $=$ \> (1) \\
$I(p) \cap M^I((x_0,\ldots,x_{n-1}))$
  \> $=$ \> (2) \\
$I(p) \cap (n\to D)$ \> $=$ \> (3) \\
$I(p)$.
\end{tabbing}
(1) Theorem~\ref{thm:atomComp},
(2) Example~\ref{ex:trivCase},
and (3) $I(p) \subseteq (n\to D)$.
\hfill$\Box$

Informally it is clear that the role of the arguments
in $p(t_0,\ldots,t_{n-1})$ is to restrict the extent of
the relation denoted by $p$. 
In Corollary~\ref{cor:free} we see that if $(t_0,\ldots,t_{n-1})$
is as unrestrictive as possible, then we get all of $I(p)$ back.
In Example~\ref{ex:diag} we see that if $(t_0,\ldots,t_{n-1})$
is as restrictive as possible, then we get back only a 
diagonal slice of $p$.

In Theorem~\ref{thm:atomComp} the right-hand side
is an expression of the intuition of the arguments
restricting the meaning of the relation denoted by the
predicate symbol.
Naively, the right-hand side should equal 
$M^I(p(t_0,\ldots,t_{n-1}))$, but this is a $C$-type relation,
whereas the right-hand side is a $P$-type relation.
The simplest fix of the discrepancy
that is at least {\sl prima facie} correct is
the composition in the left-hand side.
That a simple fix yields a correct formula is encouraging.
\section{Compositional semantics of terms}\label{sec:decTerm}

$M^I(f(t_0,\ldots,t_{n-1}))$ is determined
by Definition~\ref{def:denotT}.
Yet it is also 
determined by the denotation of its constituent $f$
(according to $I$) and by
that of its constituent $(t_0,\ldots,t_{n-1})$
(according to Definition~\ref{def:denotTT}).
The following shows the way in which
$f(t_0,\ldots,t_{n-1})$ is determined
by the denotations of its constituents.

\begin{figure}
\begin{center}
\setlength{\unitlength}{1.0cm}
\begin{picture}(13,6)
\put(2.5,5.0){\vector(1,0){7}}
\put(4.0,5.3){$M_\alpha^I(f(t_0,\ldots,t_{n-1})$}
\put(0.3,4.8){$\alpha\in(V\to D)$}
\put(0.5,4.5){\vector(1,-2){2.0}}
\put(1.0,2.5){$x\in(n\to V)$}
\put(4.5,0.8){\vector(3,2){5.5}}
\put(6.0,2.5){$I(f)\downarrow M^I(t_0,\ldots,t_{n-1})$}
\put(9.8,4.8){$b\in D$}
\put(2.8,0.3){$(x\tr \alpha)\in (n\to D)$}
\end{picture}
\end{center}
\caption{\label{fig:triangle}
Diagram to illustrate Theorem~\ref{thm:termComp},
which asserts that the triangle commutes.
}
\end{figure}
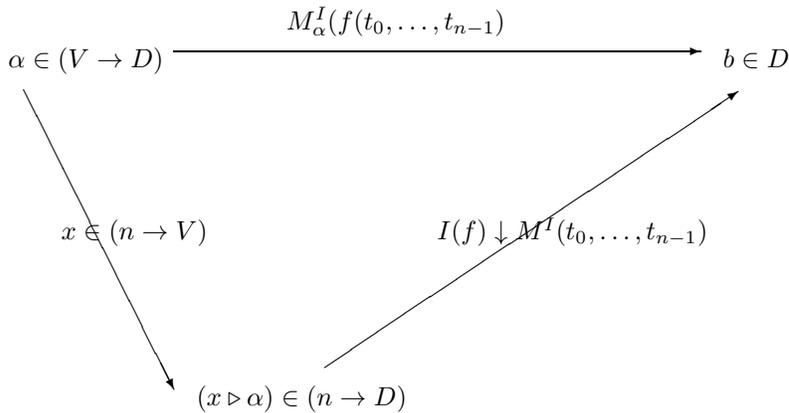

\begin{theorem}\label{thm:termComp}
Let $L$ be a signature with a function symbol $f$.
Let $I$ be an $L$-interpretation with universe of discourse $D$.
Let $t_0,\ldots,t_{n-1}$ be $L$-terms with $V$
as union of the sets of their variables.

We assert that for all $\alpha \in (V\to D)$
there exists an $x\in(n\to V)$ such that
$$
(M^I(f(t_0,\ldots,t_{n-1})))(\alpha) =
(I(f) \downarrow M^I(t_0,\ldots,t_{n-1}))(x\tr \alpha).
$$
\end{theorem}
See Figure~\ref{fig:triangle}.

\noindent
{\sl Proof.}\\
With $b=(M^I(f(t_0,\ldots,t_{n-1})))(\alpha)$ we have
\begin{tabbing}
MMMMMMMMMMMMMMMMMMMMMM\= MM\= \kill
$M^I(f(t_0,\ldots,t_{n-1}))(\alpha)$ \> $=$ \> (1)\\
$(\lambda \beta\in (V\to D).\;
   M_\beta^I(f(t_0,\ldots,t_{n-1})))(\alpha)$ \> $=$ \> (2)\\
$M_\alpha^I(f(t_0,\ldots,t_{n-1})) $ \> $=$ \> (3)\\
$(I(f))(M_\alpha^I(t_0),\ldots,M_\alpha^I(t_{n-1}))=b$.
\end{tabbing}
(1): Definition~\ref{def:denotF},
(2): beta reduction, and
(3): Definition~\ref{def:textbook}.

\vspace{2mm}
Furthermore,
\begin{tabbing}
MMMMMMMMMMMMMMMMMMMMMMMMMMMMMMMMMMMM\= MM\= \kill
$(I(f))(M_\alpha^I(t_0),\ldots,M_\alpha^I(t_{n-1}))=b$
   \> {\it iff} \> (1)\\
$\exists d\in (n\to D).\;
   (I(f))(d) = b \wedge
      d=(M_\alpha^I(t_0),\ldots,M_\alpha^I(t_{n-1}))
         \wedge d = x \tr \alpha$ \> {\it iff} \> (2)\\
$\exists d\in (n\to D).\;
   (I(f))(d) = b \wedge
      d\in M^I(t_0,\ldots,t_{n-1})
         \wedge d = x \tr \alpha$ \> {\it iff} \> (3)\\
$(I(f) \downarrow M^I(t_0,\ldots,t_{n-1}))(x\tr \alpha) = b.$
\end{tabbing}
(1): $I(f)$ is a function in $(n\to D)\to D$,
(2): Definition~\ref{def:denotTT}, and
(3): Definition~\ref{def:rangeRestr} for function restriction.
\hfill$\Box$

\section{Conclusions}

In 1933 Tarski identified first-order predicate logic \cite{tarski33}
as a language in which the concept of truth
can be defined mathematically.
The definition Tarski gave there, though mathematical,
still fell short of the standard
set by the semantics of propositional logic.
This semantics can be specified as a homeomorphism
from the syntax algebra of propositions
to the semantic algebra of truth values.
This homeomorphism is not an isolated example,
witness the following quote from \cite{janssen01}:
\begin{quotation}
{\sl
A technical description of the standard interpretation
[of compositionality] is that syntax and semantics are algebras,
and meaning assignment is a homomorphism from syntax to semantics.
This definition of compositionality is found with authors
such as Montague \cite{montague70},
Janssen \cite{janssen97},
and Hodges \cite{hodges01}.
} 
\end{quotation}

Compositional semantics for first-order predicate logic
is described in \cite{tarski52b,hmt71}.
This semantic algebra, the cylindric set algebra,
has relations as carrier.
These relations can be regarded as generalized truth values.

To us the most attractive feature of the algebraic approach
is its compositional nature:
the meaning of a composite expression
 is the result of a set-theoretically defined operation
on the meanings of it components.
In the existing work the decompositions thus treated
were conjunction, disjunction, negation and quantification.
There is the additional limitation that function symbols are absent.
In this paper we give compositional semantics
for the full language of first-order predicate logic,
including function symbols.
Moreover, we decompose atomic formulas and terms
into their constituent predicate or function symbols
and tuples of arguments.

The original Tarskian semantics specifies the conditions
under which a sentence is satisfied by an interpretation.
An interpretation assigns a relation
as meaning to each predicate symbol,
a function as meaning to each function symbol.
In the algebraic semantics of Tarski et. al.
there is no role for interpretation,
so it is not possible to decompose the meaning of an atomic formula:
there is no meaning assigned to the bare predicate symbol.
In this paper we take interpretations into account
and we give set-theoretic counterparts for conjunction,
disjunction, negation, quantification,
as well as application of predicate and function symbols
to their arguments.

It is not surprising that our semantic structure
is not the same as the one in the algebraic semantics
of Tarski et. al.
Both structures can be described in terms of two parameters:
a set $I$ of indexes and a set $D$ which is the algebraic
counterpart of the universe of discourse.
In cylindric algebra semantics the carrier consists of the
subsets of $I\to D$.
In our case it is the set of all relations of type $I'\to D$
with $I'\subseteq I$.
In Definition~\ref{def:relation} we identified the most
important operations.
In \cite{nemeti91} many algebras for quantifier logics are described.
For the purpose of this paper it does not matter
which, if any, of these matches our carrier and set of operations.

\section*{Acknowledgements}
The University of Victoria and the Natural Sciences and Engineering
Council of Canada provided facilities.

\bibliographystyle{plain}

\end{document}